\documentclass[%
reprint,
amsmath,amssymb,
aps,
pre,
floatfix,
]{revtex4-1}

\usepackage{graphicx}
\usepackage{times}
\usepackage{bm}
\usepackage{hyperref}
\usepackage{dcolumn}
\usepackage{pgf}
\usepackage{pgfplots}
\pgfplotsset{compat=1.6}

\newcommand{\ba}[1]{\bm{a}^{(#1)}}

\newcommand{\bd}[1]{\bm{d}^{(#1)}}
\newcommand{\bc}{\bm{c}}
\newcommand{\bdelta}{\bm{\delta}}
\newcommand{\bsigma}{\bm{\sigma}}

\newcommand{\bk}{\bm{k}}
\newcommand{\bu}{\bm{u}}
\newcommand{\bv}{\bm{v}}
\newcommand{\bx}{\bm{x}}
\newcommand{\bxi}{\bm{\xi}}
\newcommand{\bH}[1]{{\cal H}^{(#1)}}
\newcommand{\bD}{{\cal D}}
\newcommand{\f}[1]{f^{(#1)}}
\newcommand{\pp}[2]{\frac{\partial #1}{\partial #2}}
\newcommand{\dd}[2]{\frac{d #1}{d #2}}

\newcommand{\pe}{\textsl{Pe}}
\newcommand{\re}{\textsl{Re}}
\newcommand{\pr}{\textsl{Pr}}

\begin{document}

\title{Rotation symmetry of the multiple-relaxation-time collision model}

\author{Xuhui Li}
\author{Xiaowen Shan}
\email{shanxw@sustech.edu.cn}
\affiliation{
	Guangdong Provincial Key Laboratory of Turbulence Research and Applications\\
	Shenzhen Key Laboratory of Complex Aerospace Flows\\
	Department of Mechanics and Aerospace Engineering,
  Southern University of Science and Technology, Shenzhen, 518055, China}

\begin{abstract}
In the Hermite-expansion-based multiple-relaxation-time lattice Boltzmann (LB)
model [Shan \& Chen, \textit{Int. J. Mod. Phys. C}, \textbf{18}, 635, (2007)], a
separate relaxation time is assigned to each of the tensorial moments of the
collision term.  Here we point out that to allow maximum flexibility while
preserving the rotational symmetry of the relaxation physics, separate
relaxation times can be assigned to the components of a tensor corresponding to
its irreducible representation of SO(3) but not any finer.  By decomposing the
second moment in the LB model for polyatomic gases [Nie, Shan \& Chen,
\textit{Phys. Rev. E} \textbf{77}, 035701, (2008)], a model with decoupled shear
and bulk viscosity is constructed.  Hydrodynamic equation of the model is
obtained \textit{via} Chapman-Enskog calculation and verified by numerical
simulation.
\end{abstract}

\maketitle

\section{Introduction}

A well-known deficiency of the lattice Boltzmann-BGK (LBGK)
approach~\cite{Chen1998a} is its simplistic single-relaxation-time (SRT)
collision operator~\cite{Bhatnagar1954} adopted from continuum kinetic theory.
It relaxes all moments of the distribution function with a single rate,
resulting in the un-physical artifact that the thermal diffusivity and viscosity
are always the same.  In continuum kinetic theory a couple of models were
proposed~\cite{Holway1966,Shakhov1972} to decouple the thermal diffusivity from
viscosity by modifying the equilibrium distribution. In the context of LBGK, a
multiple-relaxation-time (MRT) model~\cite{d'Humieres1992,D'Humieres2002} was
suggested to assign separate relaxation rates to the eigenvectors of the
collision matrix in the space of discrete velocities.  The eigenvectors
represent the hydrodynamic moments that one is concerned with. Nevertheless, as
the underlying lattices used therein are insufficient to accurately represent
the third moments and beyond, the Fourier equation of heat transfer is beyond
the reach, leaving an adjustable Prandtl number unattainable.  However, the
numerical stability was indeed drastically improved, essentially due to the
trimming of the moments not fully supported by the underlying
lattice~\cite{Latt2005}.

The idea of MRT was later applied to the high-order LB which employs lattices
accurate enough to recover the full Navier-Stokes-Fourier
equations~\cite{Shan2007}.  The collision term is expanded in terms of the
tensorial Hermite polynomials~\cite{Grad1949} of which, each term is assigned an
independent relaxation time. The thermal diffusivity is decoupled from the
viscosity as they are dictated by the third and second moments respectively. A
remaining issue is that thermal diffusion is not Galilean invariant when the
thermal diffusivity is set to be different from the viscosity.  This abnormality
was later removed by explicitly correcting the third-order moment~\cite{Chen2014b},
or more systematically, by carrying out the Hermite expansion in the reference
frame moving with the fluid~\cite{Li2019,Shan2019}, or relaxing the
\textit{central moments} directly.

As each of the tensorial moments consists of multiple components, a question
arises as what the minimum unit is that can be assigned a separate relaxation
time. As a fundamental requirement, the physics of relaxation should be
independent of the coordinate system and invariant under spatial rotation.
Therefore the tensor components that can be assigned independent relaxation
times must form an \textit{irreducible representation} of the rotation group
SO(3)~\cite{Zee2016}.  For the second moment which is a symmetric rank-2 tensor
by definition, these irreducible components consist of a traceless symmetric
tensor and a unit tensor multiplied by a scalar. Hence, the linear relaxation of
the second moment can accommodate at most two rates, in analogues to the two
coefficients in the constitutive relation that give rise to the shear and bulk
viscosities~\cite{Landau1987}.

At the molecular level, bulk viscosity~\cite{Emanuel1990,Chikitkin2015} stems
from the finite time it takes for energy to equilibrate among degrees of freedom
of molecular motion. Although ignored in most cases, once activated in some scenarios 
the effects of the bulk viscosity can be significant as its value can be several thousand times of the
shear viscosity in gases such as CO$_2$, which is the main ingredient of Venus and Mars
atmospheres.  The original BGK equation models the Boltzmann equation of a
monatomic gas and hence has zero bulk viscosity.  The early athermal LBGK model,
which does not enforce the conservation of energy, exhibits an artificial
non-zero bulk viscosity which can be corrected explicitly~\cite{Dellar2001}.
Once a sufficiently accurate lattice~\cite{Shan2010c,Shan2016} is adopted so
that the full Navier-Stokes equations is recovered, the bulk viscosity does
vanish. A LBGK model for gases with internal degrees of freedom~\cite{Nie2008}
was previously proposed to model the equilibration of energy among degrees of
freedom by a pair of distributions in the reduced phase space. This model,
essentially the same as the approach used to reduce the BGK equation to lower
dimensions~\cite{Chu1965}, employs a single relaxation time and has an
adjustable heat capacity ratio, $\gamma$, and a bulk viscosity which has a fixed
ratio to the shear viscosity once $\gamma$ is fixed.

In the present work, based on the mathematical fact on the minimum rotational
invariant components that a tensor can be decomposed into, we construct a
Hermite-expansion-based MRT collision model with maximum number of relaxation
times. Particularly the second moment in the polyatomic model~\cite{Nie2008} is
decomposed into the two parts corresponding to shear and bulk viscosities to
arrive at a model with adjustable bulk-to-shear viscosity ratio 
independent of the specific heat ratio $\gamma$.
Instead of predicting experimental measurements, our goal is to derive a model
that allows the bulk viscosity freely adjusted without violating fundamental
principles.  The rest of the paper is organized as the following.  In
Sec.~\ref{sec:intro} we give the theoretical derivation.  After a brief review
of of the background, we present the tensor decomposition if Sec.~\ref{sec:rot},
followed by an introduction of the SRT polyatomic gas model in
Sec.~\ref{sec:poly}, and its extension to MRT in Sec.~\ref{sec:mrt}.  The
hydrodynamic equation of the model is derived \textit{via} Chapman-Enskog
calculation in Sec.~\ref{sec:ce}.  In Sec.~\ref{sec:numerical} numerical
verification is presented, and finally conclusions and some discussions are
given in Sec.~\ref{sec:discussion}.

\section{Theoretical derivation}\label{sec:intro}

In a previous series of papers~\cite{Shan2007,Li2019,Shan2019} we propose to
define the collision process through its action on the non-equilibrium part of
the distribution function.  Specifically we expand the distribution function in
terms of Hermite polynomials~\cite{Grad1949} as:
\begin{equation}
  \label{eq:expansion}
  f(\bxi, \bx, t) = \omega(\bm{\eta})\sum_{n=0}^\infty
  	\frac{1}{n!}\ba{n}(\bx, t):\bH{n}(\bm{\eta}),
\end{equation}
where $\bm{\eta}$ can be $\bxi$, $\bc\equiv\bxi-\bu$, or $\bv\equiv \bc /
\sqrt{\theta}$, corresponding respectively to expansions in the laboratory
frame, frame moving with the fluid, and thermally-scaled moving frame
respectively.  Here, $\bc$ is the \textit{peculiar velocity}, $\theta$ the
temperature and $\omega(\bm{\eta}) \equiv (2\pi)^{-D/2}\exp(-\eta^2/2)$ the
weight function. The expansion coefficients, $\ba{n}(\bx, t)$, are the moments
of the distribution function, or their combinations, in the various frames.  The
whole set of $\ba{n}$ completely and uniquely specifies $f(\bxi, \bx, t)$ and
\textit{vice versa}. In case $\bc$ or $\bv$ is used, the \textit{binomial
	transform} and a further scaling can be used to transform $\ba{n}$ back to the
laboratory frame where they can be exactly represented by a set of fixed
discrete velocities~\cite{Shan2006b}.

Now consider the collision operator $\Omega(f)$ which represents the change to
the distribution due to the local collision process.  As apparently
$\Omega(\f{eq}) = 0$, for convenience, we denote the non-equilibrium part of the
distribution by $\f{neq}\equiv f - \f{eq}$ and redefine $\Omega$ as a functional
of $\f{neq}$ such that $\Omega(0) = 0$.  For instance, the well-known BGK
collision operator is simply $\Omega(f) = -\omega f$ where $\omega\equiv 1/\tau$
is the collision frequency and $\tau$ the collision time.  Let the expansion
coefficients of $\f{neq}$ and $\Omega(\f{neq})$ in terms of $\bH{n}(\bv)$ be
$\bd{n}_1$ and $\bd{n}_\Omega$ respectively.  Note that if the construction of
$\f{eq}$ guarantees the conservation of mass, momentum and energy, we have
$\bd{0}_1 = 0$, $\bd{1}_1 = \bm{0}$ and $\bd{2}_1$ is traceless. The collision
operator can be specified \textit{via} the expansion coefficients. The previous
MRT model is defined by~\cite{Li2019}:
\begin{equation}
  \bd{n}_\Omega = -\omega_n\bd{n}_1,\quad n = 2, 3, \cdots,
\end{equation}
which gives each of the Hermite terms a separate relaxation frequency,
$\omega_n$.

\subsection{Rotational symmetry of rank-2 tensor}\label{sec:rot}

A well-known conclusion of group theory~\cite{Zee2016} is that the 9-dimensional
(9-d) representation of the rotation group SO(3) furnished by a rank-2 tensor
can be decomposed into a 5-d space of traceless symmetric tensor, a 3-d space of
anti-symmetric tensor, and a 1-d space of unit tensor, commonly noted as
$5\oplus 3\oplus 1$, each of which is closed under the transform of SO(3).
Hence, the second tensorial moment, which is symmetric by definition, can be
decomposed into the spaces of a traceless symmetric tensor and a unit tensor
($5\oplus 1$), both can be relaxed separately without breaking rotational
symmetry.

To further illustrate, let us define two operators, $\bar{\bm{a}}$ and
$\tilde{\bm{a}}$, which respectively take the trace and traceless symmetric
component of the rank-2 tensor, $\bm{a}$. Assuming Einstein summation, in
component form we define:
\begin{equation}
\bar{a}\equiv a_{ii},\quad\mbox{and}\quad
\tilde{a}_{ij}\equiv\frac 12\left(a_{ij} + a_{ji}\right)
-\frac{\bar{a}}D\delta_{ij}.
\end{equation}
Any symmetric rank-2 tensor can be decomposed as:
\begin{equation}
  \label{eq:decomp}
  \bm{a} = \frac{\bar{a}}D\bdelta + \tilde{\bm{a}}.
\end{equation}
Particularly, the second-order term in a Hermite expansion can be decomposed
into two parts that are orthogonal under spatial rotation:
\begin{equation}
  \label{eq:5}
  \bm{a}:\bH{2}(\bv) = \tilde{\bm{a}}:\bH{2}(\bv)+\frac{\bar{a}}{D}\bdelta:\bH{2}(\bv).
\end{equation}
The action of the most general rotationally symmetric linear relaxation operator
on such a functional can be defined as:
\begin{equation}
  \Omega\left[\bm{a}:\bH{2}(\bv)\right] = -\left[\frac 1{\tau_1}\tilde{\bm{a}} 
  + \frac 1{\tau_2}\frac{\bar{a}}{D}\bdelta \right] :\bH{2}(\bv),
\end{equation}
with two independent relaxation times, $\tau_1$ and $\tau_2$.

We note that for the two tensor contractions on the right-hand-side of
Eq.~(\ref{eq:5}), only the traceless component and the trace of $\bH{n}(\bv)$
have contributions.  Hence the same term can take several equivalent forms,
\textit{e.g.}:
\begin{equation}
  \bm{a}:\bH{2}(\bv) = \tilde{\bm{a}}:\bv\bv + \frac{\bar{a}}D\left(v^2-D\right).
\end{equation}

\subsection{BGK model with internal degrees of freedom}\label{sec:poly}

For a monatomic gas in which the translational kinetic energy is conserved,
$\bd{2}_1$ is traceless so that $\overline{\bd{2}_1}= 0$. Only a single
relaxation time can exist. Previously a lattice BGK model was proposed for gases
with internal degrees of freedom~\cite{Nie2008}. The approach is essentially the
same as the one developed to reduce spatial dimensionality~\cite{Chu1965}. In
this approach, a polyatomic gas is described by a pair of distribution functions
in the \textit{reduced} phase space, $(\bxi, \bx)$, obeying the following BGK
equations:
\begin{subequations}
  \label{eq:polyatomic}
  \begin{eqnarray}
	\label{eq:g1}
	\pp{g}t + \bxi\cdot\nabla g &=& \Omega_g\equiv -\frac 1\tau\left[g-g^{(eq)}\right],\\
	\pp{h}t + \bxi\cdot\nabla h &=& \Omega_h\equiv -\frac 1\tau\left[h-g^{(eq)}\theta\right],
  \end{eqnarray}
\end{subequations}
where $\Omega_g$ and $\Omega_h$ are the collision operators for $g$ and $h$, and
\begin{equation}
  \label{eq:geq}
  g^{(eq)} = \frac\rho{\left(2\pi\theta\right)^{D/2}}\exp
  \left(-\frac{c^2}{2\theta}\right) = \frac\rho{\theta^{D/2}}\omega(\bv),
\end{equation}
is the Maxwellian in the reduced phase space. Here $\rho$ is the density,
$c\equiv|\bc|$, and $\theta$ the temperature which is related to the
\textit{energy density} per mass, $\epsilon$, by:
\begin{equation}
  \label{eq:epsilon}
  \epsilon = \frac 12\left(D+S\right)\theta.
\end{equation}
where $S$ is the number of internal degrees of freedom.  The variables $\rho$,
$\bu$, and $\epsilon$, are moments of the two distributions:
\begin{subequations}
  \begin{eqnarray}
	\rho &=& \int gd\bxi,\\
	\rho\bu &=& \int g\bxi d\bxi,\\
	\label{eq:energy}
	\rho\epsilon &=& \frac 12\int gc^2d\bxi + \frac S2\int hd\bxi.
  \end{eqnarray}
\end{subequations}

Although Eq.~(\ref{eq:g1}) appears to be identical to the BGK equation for
monatomic gases, a critical difference lies in the calculation of $g^{(eq)}$
where $\theta$ is now given by Eqs.~(\ref{eq:epsilon}) and (\ref{eq:energy})
which couple $g$ and $h$. This coupling reflects energy transfer between the
translational and internal degrees of freedom. As the translational energy is
not conserved, $g^{(neq)}\equiv g-g^{(eq)}$ can have a non-vanishing trace.
Nevertheless, the \textit{total} energy is still conserved as from
Eqs.~\eqref{eq:energy} and \eqref{eq:epsilon}, we have:
\begin{equation}
  \label{eq:10}
  \int gc^2d\bxi + S\int hd\bxi = (D+S)\rho\theta.
\end{equation}
The definition of $g^{(eq)}$ of Eq.~\eqref{eq:geq} gives:
\begin{equation}
  \int g^{(eq)}c^2d\bxi = D\rho\theta,\quad\mbox{and}\quad
  \int g^{(eq)} d\bxi = \rho.
\end{equation}
The above two equations lead to:
\begin{equation}
  \label{eq:eneq}
  \int g^{(neq)}c^2d\bxi + S\int h^{(neq)}d\bxi = 0.
\end{equation}
Together with the fact that the two relaxation times in
Eqs.~\eqref{eq:polyatomic} are identical, it ensures that:
\begin{equation}
  \label{eq:conserve}
  \int\left(\Omega_g c^2 + S\Omega_h\right)d\bxi = 0,
\end{equation}
namely, $\rho\epsilon$, as defined by Eq.~(\ref{eq:energy}), is conserved by the
collision operator.  Obviously mass and momentum are also conserved as:
\begin{equation}
  \label{eq:conserve1}
  \int\Omega_gd\bxi = 0,\quad\mbox{and}\quad
  \int\Omega_g\bxi d\bxi = 0.
\end{equation}

\subsection{MRT extension}\label{sec:mrt}

Now let $\bd{n}_g$ and $\bd{n}_h$ be respectively the Hermite expansion
coefficients of $g^{(neq)}$ and $h^{(neq)}\equiv h-g^{(eq)}\theta$,
\textit{i.e.}:
\begin{subequations}
  \label{eq:neq}
\begin{eqnarray}
  \label{eq:gneq}
  g^{(neq)} &=& \omega(\bv)\sum_{n=2}^\infty\frac 1{n!}\bd{n}_g:\bH{n}(\bv),\\
  \label{eq:hneq}
  h^{(neq)} &=& \omega(\bv)\sum_{n=0}^\infty\frac 1{n!}\bd{n}_h:\bH{n}(\bv).
\end{eqnarray}
\end{subequations}
Due to the conservation of mass and momentum, $\bd{0}_g$ and $\bd{1}_g$ vanish.
The leading coefficient of $g^{(neq)}$ is:
\begin{eqnarray}
  \lefteqn{\bd{2}_g = \int \left[g-g^{(eq)}\right]\bH{2}(\bv)d\bv} \nonumber\\
  &=& \int(\bv\bv-\bdelta)gd\bv 
  = \theta^{-\frac{D+2}2}\left[\int g\bc\bc d\bc - \rho\theta\bdelta\right],
\end{eqnarray}
where the fact $\int g^{(eq)}\bH{2}(\bv)d\bv = 0$ is used. The trace and
traceless component of $\bd{2}_g$ can be computed as:
\begin{subequations}
  \begin{eqnarray}
    \label{eq:16a}
	\bar{d}^{(2)}_g &=& \theta^{-\frac{D+2}2}
		\left[\int gc^2 d\bc - D\rho\theta\right],\\
	\tilde{\bm{d}}^{(2)}_g &=& \theta^{-\frac{D+2}2}
		\int g\left[\bc\bc - \frac{c^2}D\bdelta\right] d\bc.
  \end{eqnarray}
\end{subequations}
Similarly the leading coefficient of $h^{(neq)}$ is:
\begin{equation}
  d^{(0)}_h = \int\left[h - \frac{\rho\theta}{\theta^{D/2}}\omega(\bv)\right]d\bv
  = \theta^{-\frac D2}\left[\int hd\bc - \rho\theta\right].
\end{equation}
Using Eqs.~(\ref{eq:epsilon}) and (\ref{eq:energy}), $d^{(0)}_h$ is related to
$\bar{d}^{(2)}_g$ by:
\begin{equation}
  \theta\bar{d}^{(2)}_g + Sd_h^{(0)} = 0.
\end{equation}

Using Eq.~(\ref{eq:decomp}), the leading term in Eq.~(\ref{eq:gneq}) can be
decomposed to have:
\begin{equation}
\frac{g^{(neq)}}{\omega(\bv)} = \frac{1}{2!}\left[\tilde{\bm{d}}^{(2)}_g
+ \frac{\bar{d}^{(2)}_g}{D}\bdelta\right]:\bH{2}
+ \frac{\bm{d}^{(3)}_g:\bH{3}}{3!}+\cdots
\end{equation}
A corresponding MRT collision model can then be devised as:
\begin{subequations}
	\label{eq:mrt}
	\begin{eqnarray}
	\label{eq:og}
	-\frac{\Omega_g}{\omega(\bv)} &=& \frac{1}{2}\left[
	  \frac{\tilde{\bm{d}}^{(2)}_g}{\tau_{21}}
	+ \frac{\bar{d}^{(2)}_g\bdelta}{\tau_{22}D}
	\right]:\bH{2} +\frac{\bd{3}_g:\bH{3}}{3!\tau_3}+\cdots,\\
	\label{eq:oh}
	-\frac{\Omega_h}{\omega(\bv)} &=& \frac{d_h^{(0)}}{\tau_0}
	+\frac{\bm{d}^{(1)}_h:\bH{1}}{\tau_1}+\cdots,
	\end{eqnarray}
\end{subequations}
where, $\tau_{21}$, $\tau_{22}$, $\tau_3$, $\tau_0$ and $\tau_1$ are independent
relaxation times. We now show that the energy conservation of
Eq.~(\ref{eq:conserve}) demands that $\tau_0 = \tau_{22}$. For translational
energy we note that:
\begin{eqnarray}
  \lefteqn{\int\Omega_g c^2d\bxi = \theta^{\frac{D+2}2}\mbox{Tr}
  	\int\Omega_g\bv\bv d\bv}\nonumber\\
	&=& \theta^{\frac{D+2}2}\mbox{Tr}
	\int\Omega_g\left[\bH{2}(\bv) + \bdelta\bH{0}(\bv)\right]d\bv.
\end{eqnarray}
Using the orthogonal relations and the fact that $\mbox{Tr}
(\tilde{\bm{d}}^{(2)}_g) = 0$, we have:
\begin{equation}
  \int\Omega_g c^2d\bxi = -\frac{\theta^{\frac{D+2}2}\bar{d}^{(2)}_g}{\tau_{22}}
  =-\frac{1}{\tau_{22}}\left[\int gc^2 d\bc - D\rho\theta\right],
\end{equation}
and similarly:
\begin{equation}
  \int\Omega_hd\bxi = -\frac{\theta^{\frac D2}d^{(0)}_h}{\tau_0}
   = -\frac{1}{\tau_0}\left[\int hd\bc - \rho\theta\right].
\end{equation}
Noticing Eq.~(\ref{eq:10}), in order for Eqs.~(\ref{eq:mrt}) to satisfy
Eq.~(\ref{eq:conserve}), we must have $\tau_0 = \tau_{22}$.

\subsection{Hydrodynamic equations}\label{sec:ce}

We now derive the hydrodynamic equations of Eqs.~(\ref{eq:polyatomic}) and
(\ref{eq:mrt}).  By taking the moments in Eqs.~(\ref{eq:conserve1}) and
(\ref{eq:conserve}) of Eqs.~(\ref{eq:polyatomic}), we have the conservation
equations:
\begin{subequations}
  \label{eq:cons1}
  \begin{eqnarray}
  \dd{\rho}t &+& \rho\nabla\cdot\bu = 0,\\
  \rho\dd{\bu}t &+& \nabla\cdot\bm{P} = 0,\\
  \rho\dd\epsilon t &+& \nabla\bu:\bm{P} + \nabla\cdot\bm{q} = 0,
  \end{eqnarray}
\end{subequations}
where $d/dt\equiv\partial/\partial t +\bu\cdot\nabla$ is the \textit{material
	derivative}, and
\begin{equation}
  \label{eq:pq}
  \bm{P}\equiv\int g\bc\bc d\bc,\quad\mbox{and}\quad
  \bm{q}\equiv\frac 12\int (gc^2 + Sh)\bc d\bc,
\end{equation}
are the \textit{pressure tensor} and \textit{energy flux} respectively.  Except
for the last term in the definition of $\bm{q}$, all are the same as in a
monatomic gas. The \textit{hydrostatic pressure}:
\begin{equation}
  p\equiv\frac{P_{ii}}D = \frac 1D\int gc^2d\bc,
\end{equation}
is defined as the average of the normal components of $\bm{P}$ and the
\textit{deviatoric stress}, $\bsigma\equiv -(\bm{P} - p\bdelta)$, is the
negative of the traceless part of $\bm{P}$. In particular, if $g$ and $h$ are
expanded in terms of Hermite polynomials $\bH{n}(\bv)$, we have:
\begin{eqnarray}
  \bm{P} &=& \theta^{\frac{D+2}2}\int g\left[\bdelta + \bH{2}(\bv)\right]d\bv\nonumber\\
  &=& \theta^{\frac{D+2}2}\left[d^{(0)}\bdelta + \bm{d}^{(2)}\right].
\end{eqnarray}
On decomposing $\bd{2}$ according to Eq.~(\ref{eq:decomp}), we have:
\begin{equation}
  p = \theta^{\frac{D+2}2}\left[d^{(0)} + \frac{\bar{d}^{(2)}}{D}\right],
  \quad\mbox{and}\quad
  \bsigma = -\theta^{\frac{D+2}2}\tilde{\bm{d}}^{(2)}.
\end{equation}

As the zero-th approximation, taking $g$ and $h$ as their equilibria, $g^{(eq)}$
and $g^{(eq)}\theta$, we have the ideal gas equation of state:
\begin{equation}
p = \rho\theta,
\end{equation}
and
\begin{equation}
\bsigma = \bm{0},\quad\mbox{and}\quad \bm{q} = \bm{0},
\end{equation}
which yield Euler's equations when plugged into Eqs.~(\ref{eq:cons1}). The first
Chapman-Enskog approximation~\cite{Huang1987} amounts to substituting $f =
\f{eq} + \f{1} + \cdots$ into the Boltzmann-BGK equation and keeping the leading
terms on both sides to yield:
\begin{equation}
  \f{1} \cong -\tau\left(\pp{}t+\bxi\cdot\nabla\right)\f{eq}.
\end{equation}
Using Euler's equation to convert the time derivatives to spatial ones, $\f{1}$
can be written in terms of the hydrodynamic variables and their spatial
derivatives. On substituting into Eq.~(\ref{eq:pq}), we obtain $\bm{P}^{(1)}$
and $\bm{q}^{(1)}$ and in turn the Navier-Stokes equations when plugged into
Eqs.~(\ref{eq:cons1}).

It was noted~\cite{Shan2019} that in this procedure, $\bm{P}^{(1)}$ and
$\bm{q}^{(1)}$ can be obtained by taking the corresponding moments of the BGK
equation directly without computing $\f{1}$ explicitly. We now apply the same
procedure to Eqs.~(\ref{eq:polyatomic}) and (\ref{eq:mrt}). Define the
deferential operator for brevity:
\begin{equation}
\bD\equiv\pp{}t + \bxi\cdot\nabla = \dd{{}}t + \bc\cdot\nabla,
\end{equation}
the first Chapman-Enskog approximation can be written as:
\begin{widetext}
\begin{subequations}
\begin{eqnarray}
  \label{eq:ce}
  \bD g^{(eq)} &=& -\omega(\bv)\left[
  \frac 12\left(\frac{\tilde{\bm{d}}^{(2)}_g}{\tau_{21}}
   + \frac{\bar{d}^{(2)}_g\bdelta}{\tau_{22}D} \right):\bH{2}
   + \frac{\bd{3}_g:\bH{3}}{3!\tau_3}
   + \cdots\right],\\
  \label{eq:ce-h}
  \bD (g^{(eq)}\theta) &=& -\omega(\bv)\left[\frac{d_h^{(0)}}{\tau_{22}}
   + \frac{\tilde{\bm{d}}^{(1)}_h:\bH{1}}{\tau_1}
   + \cdots\right].
\end{eqnarray}
\end{subequations}
Following the standard procedure~\cite{Huang1987}, the left-hand-side can be
computed as:
\begin{subequations}
\begin{eqnarray}
\label{eq:40a}
\bD g^{(eq)} &=& \frac 1\theta\left[\left(\bc\bc - \frac{c^2\bdelta}D\right):\nabla\bu
 + \left(\frac{c^2}{2\theta}-\frac{D+2}2\right)\bc\cdot\nabla\theta
 + \frac{S}{D+S}\left(\frac{c^2}D-\theta\right)\nabla\cdot\bu
 \right]g^{(eq)},\\
\bD (g^{(eq)}\theta) &=& \left[\left(\bc\bc - \frac{c^2\bdelta}D\right):\nabla\bu
+ \left(\frac{c^2}{2\theta}-\frac{D}2\right)\bc\cdot\nabla\theta
+ \frac{S}{D+S}\left[\frac{c^2}D-\left(1+\frac 2S\right)\theta\right]\nabla\cdot\bu
\right]g^{(eq)},
\end{eqnarray}
\end{subequations}
\end{widetext}
where the first two terms in the brackets of the first equation are the same as
in the monatomic case, while the third one is due to the internal degrees of
freedom.
Now taking the moment $\int\cdot\bc\bc d\bc$ of Eq.~(\ref{eq:ce}), we have:
\begin{eqnarray}
  &2\rho\theta&\left[\bm{\left(\Lambda}-\frac{\bdelta\nabla\cdot\bu}D\right)
   + \frac{S\bdelta\nabla\cdot\bu}{D(D+S)}\right]\nonumber\\
   &=&-\theta^{\frac{D+2}2}\left[\frac{\tilde{\bm{d}}_g^{(2)}}{\tau_{21}}
   + \frac{\bar{d}_g^{(2)}\bdelta}{D\tau_{22}}\right].
\end{eqnarray}
where $\bm{\Lambda}\equiv\frac 12\left[\nabla\bu +
\left(\nabla\bu\right)^T\right]$ is the \textit{strain rate}. Matching up the
trace and traceless part on both sides, we have the correction to the
hydrostatic pressure and deviatoric stress as:
\begin{subequations}
\begin{eqnarray}
  p^{(1)} &=& \theta^{\frac{D+2}{2}} \frac{\bar{d}_g^{(2)}}{D}
   = -\frac{2S}{D(D+S)}\tau_{22}\rho\theta\nabla\cdot\bu,\\
  \bsigma^{(1)} &=& -\theta^{\frac{D+2}{2}}\tilde{\bm{d}}_g^{(2)} = 
   2\tau_{21}\rho\theta\bm{\left(\Lambda}-\frac{\bdelta\nabla\cdot\bu}D\right),
\end{eqnarray}
\end{subequations}
corresponding to the kinematic shear viscosity, $\nu$, and bulk viscosity,
$\nu_b$:
\begin{equation}
  \nu = \tau_{21}\rho\theta,\quad\mbox{and}\quad
  \nu_b = \frac{2S}{D(D+S)}\tau_{22}\rho\theta.
\end{equation}
Similarly taking the moments $\int\cdot c^2\bc d\bc$ and $\int\cdot\bc d\bc$ of
Eqs.~(\ref{eq:ce}) and (\ref{eq:ce-h}) respectively yields:
\begin{subequations}
\begin{eqnarray}
  \theta^{\frac{D+3}2}\bm{d}_g^{(3)} &=& -\tau_3(D+2)\rho\theta\nabla\theta,\\
  \theta^{\frac{D+1}2}\bm{d}_h^{(1)} &=& -\tau_1\rho\theta\nabla\theta.
\end{eqnarray}
\end{subequations}
The first-order correction to the heat flux is then:
\begin{equation}
  \bm{q}^{(1)} = -\frac{(D+2)\tau_3 + S\tau_1}2\rho\theta\nabla\theta,
\end{equation}
corresponding to a \textit{heat conductivity} of:
\begin{equation}
  \lambda = \frac{(D+2)\tau_3 + S\tau_1}2\rho\theta.
  \label{eq:lambda}
\end{equation}
Noting that $c_p = (D+S+2)/2$, the \textit{thermal diffusivity} is:
\begin{equation}
  \kappa = \frac{(D+2)\tau_3 + S\tau_1}{D+S+2}\rho\theta,
\end{equation}
which reduces to $\tau\rho\theta$ if $S=0$ as for monatomic gases, or $\tau_1 =
\tau_3 = \tau$ as for the single-relaxation-time polyatomic
model~\cite{Nie2008}.

\section{Numerical versification}\label{sec:numerical}

To verify the model, the transport coefficients are measured form the dynamics
of the linear hydrodynamic modes in an one-dimensional periodic setup and
compared with their theoretical values. The case setup has been extensively
discussed previously~\cite{Shan2007,Li2011a,Shan2019}.  Here we briefly
summarize the analytical results.  Consider the monochromatic plane wave
perturbation:
\begin{equation}
\left(
\begin{array}{c}
\rho \\ 
\bu \\
\theta \\
\end{array}
\right)
=
\left(
\begin{array}{c}
\rho_0 \\ 
\bu_0 \\
\theta_0 \\
\end{array}
\right)
+
\left(
\begin{array}{c}
\bar{\rho} \\ 
\bar{\bu} \\
\bar{\theta} \\
\end{array}
\right)
e^{\omega t + i\bk \cdot (\bm{x}-\bu_0 t) }
\end{equation}
where the subscript $_0$ denotes the base flow and $\bar{\rho}$, $\bar{\bu}$ and
$\bar{\theta}$ are the perturbation amplitudes. Both the base state and the
perturbation amplitudes are homogeneous and constant. $\bm{k}$ and $\omega$
respectively are the \textit{wave vector} and \textit{angular frequency} of the
plane wave. Decomposing the velocity into components parallel and perpendicular
to the wave vector and substituting into Eqs.~(\ref{eq:cons1}), we obtain an
eigen system in the space of $(\bar{\rho}, \bar{u}_\parallel, \bar{\theta},
\bar{u}_\perp)^T$ from which the dimensionless dispersion relations of four
linear modes can be obtained:
\begin{subequations}
	\label{eq:dr}
	\begin{eqnarray}
	-\frac{\omega_v}{c_sk} &=& \frac 1{\re},\\
	-\frac{\omega_t}{c_sk} &=& \frac 1{\pe} + \frac{(\gamma-1)\lambda}{\pe^3}
	+ \mathcal{O}\left(\frac{1}{\pe^5}\right),\\
	\label{eq:dr-3}
	-\frac{\omega_\pm}{c_sk} &=& \frac{\gamma -\lambda}{2\pe}
	-\frac{(\gamma-1)\lambda}{2\pe^3}
	+\mathcal{O}\left(\frac{1}{\pe^5}\right)\nonumber\\
	&\pm& i\left[1-\frac{(\gamma +\lambda)^2 - 4\lambda}{8\pe^2}
	+ \mathcal{O}\left(\frac{1}{\pe^4}\right)\right],
	\end{eqnarray}
\end{subequations}
where $\omega_v$, $\omega_t$, and $\omega_\pm$ are the angular frequencies of
the viscous, thermal and two acoustic modes, $c_s\equiv\sqrt{\gamma\theta_0}$ a
{\em characteristic} speed of sound, $k\equiv |\bm{k}|$ the \textit{wave
	number}, $\re\equiv c_s/\nu k$, $\pe\equiv c_s/\kappa k$ and
$\pr\equiv\nu/\kappa$ the \textit{acoustic} Reynolds, P\'eclet and Prandtl
numbers, and
\begin{equation}
	\lambda\equiv 1 + \left(\gamma -3- \frac{\nu_b}{\nu}\right)\pr
\end{equation}
a constant defined for brevity which is the only place where bulk viscosity
affects the dispersion relations.  While the viscous mode is independent from
the other three and its dispersion relation is exact, the dispersion relations
of the thermal and acoustic modes are solutions of a cubic characteristic
equation and only their asymptotic form at large-\pe\ limit are given.  Up to
the order of $\mathcal{O}(\pe^{-2})$, the decay rates of the viscous and thermal
modes are:
\begin{equation}
  \omega_v = -\nu k^2,\quad\mbox{and}\quad
  \omega_t \cong -\kappa k^2\left[1 + \frac{(\gamma-1)\lambda}{\pe^2}\right].
\end{equation}
The \textit{sound attenuation rate} is a weighted sum of the shear viscosity,
bulk viscosity and thermal diffusivity:
\begin{equation}
  \alpha\cong - k^2\left[a\kappa + (1-a)\nu + \frac{\nu_b}2\right]
	\left[1 + \frac{(\gamma-1)\lambda}{(\gamma - \lambda)\pe^2}\right],
\end{equation}
where $a = (\gamma-1)/2$.  The speed of sound is also corrected by the
dissipation rates as:
\begin{equation}
  \sqrt{\gamma\theta_0}\left[1-\frac{(\gamma +\lambda)^2 - 4\lambda}{8\pe^2}\right].
\end{equation}
While the effects of the bulk viscosity on sound speed and decay rate of the
thermal mode is in the order of $\mathcal{O}(\pe^{-2})$, its effect on sound
attenuation is in the leading order. As a verification of the bulk viscosity, we
numerically measure the sound attenuation rate and compare with the theoretical
value.

The simulation is performed on a 3D periodic lattice of the dimension $L_x\times
L_y\times L_z$ using the minimum 9-th order $E_{3,103}^9$
quadrature~\cite{Shan2016} capable of representing 4-th moments exactly. All
simulations were conducted with $L_x = 256$, $L_y = 5$, $L_z = 5$. The initial
perturbations of all physical quantities are spatially sinusoidal wave in the
form of $\sin\frac{2\pi n_xx}{L_x} \sin\frac{2\pi n_yy}{L_y} \sin\frac{2\pi
	n_zz}{L_z}$, where the integer vector $(x, y, z)$ is the lattice coordinates,
and $n_x=1$, $n_y=n_z=0$ the wave numbers.  The wave vector is $\bm{k} =
(2\pi/c)\left(n_x/L_x, n_y/L_y, n_z/L_z\right)$ where $c$ is the lattice
constant. The dynamics of any of the three variables, $\bar{\rho}$,
$\bar{u}_\parallel$ or $\bar{\theta}$, is the superposition of the thermal and
acoustic modes.  To study the acoustic mode, we set the initial perturbation
amplitudes to the eigen state of a standing wave which is asymptotically
$(\bar{\rho}, \bar{u}_\parallel, \bar{\theta}) = (\rho_0, 0,
\theta_0(\gamma-1))$. Alternatively, as the acoustic modes are isentropic, we
can also set $\rho = \rho_0 + \rho'\sin\bm{k}\cdot\bm{x}$ and initialize
$\theta$ using the \textit{isentropic invariant} $\theta\rho^{1-\gamma} =
\theta_0\rho_0^{1-\gamma}$.  The sound attenuation rate, $\alpha$, and
frequency, $\omega$, are measured by fitting the pressure fluctuation with the
model $a_s e^{-\alpha t}\sin(\omega t + \phi)$.

\begin{figure}[!htbp]
	\centering
	\begin{tikzpicture}
	\begin{axis}[
		xlabel={Time step},
		ylabel={Pressure perturbation amplitude},
		enlarge x limits=false,
		enlarge y limits=false,
		legend style={
		  at={(0., 0.5)},
		  anchor=west
		},
		legend cell align=left,
		grid=major
	]
	\addplot[no markers,densely dotted] table[x=stp, y=d100] {history.dat};
	\addplot[no markers,dashed] table[x=stp, y=d500] {history.dat};
	\addplot[no markers,dotted,green] table[x=stp, y=d1000] {history.dat};
	\addplot[no markers,dashdotted,blue] table[x=stp, y=d1500] {history.dat};
	\addplot[no markers,red] table[x=stp, y=d2000] {history.dat};
	\legend{
		{$\nu_b/\nu = 100$},
		{$\nu_b/\nu = 500$},
		{$\nu_b/\nu = 1000$},
		{$\nu_b/\nu = 1500$},
		{$\nu_b/\nu = 2000$}
	};
	\end{axis}
	\end{tikzpicture}

	\caption{Typical time history of the amplitude of pressure perturbation in
	acoustic waves with varying ratio of bulk to shear viscosities. The other
	parameters are $\tau_{21}=0.5005$, $\gamma=1.3$, $\pr=2.1$ and $\tau_1 =
	\tau_3$.  To be seen is that the sound attenuation rate significantly depends
	on the bulk viscosity.}

	\label{figure:oscillations}
\end{figure}
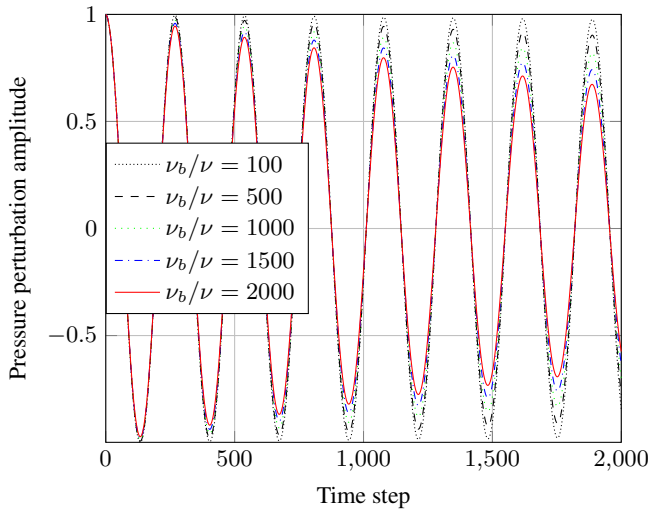

\begin{figure}[!htbp]
	\centering
	\begin{tikzpicture}
	\begin{semilogxaxis}[
	xlabel={$\nu_b/\nu$},
	ylabel={Relative Errors in sound ateenuation rate},
	ymin=-0.015,
	ymax=0.01,
	legend pos=north west,
	legend cell align=left,
	log ticks with fixed point,
	grid=major
	]
	\addplot coordinates {
		(0.05, -0.003210323)
		(0.1,  -0.003349932)
		(0.25, -0.003665123)
		(0.5,  -0.004092047)
		(1.,  -0.004668139)
		(2.5, -0.00563478)
		(5,   -0.006669458)
		(10,  -0.00907757)
	};
	\addplot coordinates {
		(50,    0.007157277)
		(100,   9.34143E-05)
		(200,   -0.003440799)
		(500,  -0.005725227)
		(1000, -0.007046266)
		(1500, -0.008273206)
		(2000, -0.009752156)
		(3000, -0.013596387)
	};
	\legend{
		{$\tau_{21} = 0.6$},
		{$\tau_{21} = 0.5005$}
	};
	\end{semilogxaxis}
	\end{tikzpicture}
	\caption{Relative error of decay rate with varying ratio of bulk viscosity to
		shear viscosity. Two cases are tested: $\tau_{21}=0.5005$ and $\tau_{21}=0.6$.
		In both cases $\gamma=1.3$, $\pr =2.1$ and $\tau_1 = \tau_3$. To be seen is
		that the relative error is generally below 1\% over a large range of
		$\nu_b/\nu$.} 
	\label{figure:relative_errors}
\end{figure}
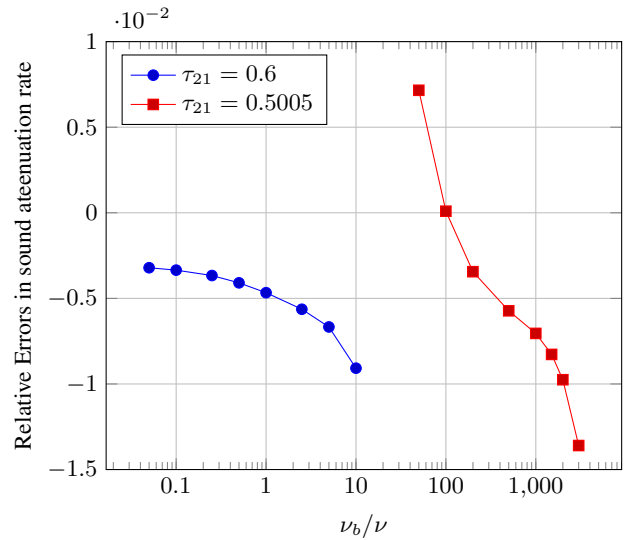

Shown in Fig.~\ref{figure:oscillations} are the time histories of the pressure
perturbation amplitude for various ratios of bulk to shear viscosities. The
shear viscosity and all other parameters are fixed. Clearly the sound
attenuation rate increases with the bulk viscosity. As shown in
Fig.~\ref{figure:relative_errors}, the relative error of sound attenuation rate
against its theoretical value is below 1\% over a wide range of the viscosity
ratio.  As a relaxation time of $\tau_{21}\lesssim 0.5005$ with small bulk viscosity could cause stability
problem, and $\tau_{22}\gtrsim 2$ corresponds to a finite Knudsen number, two
different $\tau_{21}$ is used. Also worth noting is that at least in the
continuum flow regime, as long as the total thermal conductivity remain the
same, $\tau_1$ and $\tau_3$ can be adjusted freely according the constraint
Eq.~(\ref{eq:lambda}) without causing any visible effect. The effect of $\tau_1$
and $\tau_3$ in rarefied gas flow regime remains to be investigated in the
future work.

\begin{figure}
	\centering
	\begin{tikzpicture}
	\begin{axis}[
	xlabel={Specific heat ratio, $\gamma$.},
	ylabel={Speed of sound},
	xmin=1.,
	ymin=1.,
	legend pos=north west,
	legend cell align=left,
	log ticks with fixed point,
	grid=major
	]
	\addplot [domain=1.05:1.7] {sqrt(x)};
	\addplot coordinates {
		(1.1,  1.044733484)
		(1.15, 1.068203974)
		(1.2,  1.091169945)
		(1.25, 1.113666706)
		(1.3,  1.135724299)
		(1.35, 1.157359647)
		(1.4,  1.178599769)
		(1.45, 1.199462757)
		(1.5,  1.219971485)
		(1.55, 1.24014424)
		(1.6,  1.26000931)
		(1.666667, 1.285947529)
	};
	\legend{
		{Analytical ($\sqrt{\gamma\theta_0}$)},
		{Measured}
	};
	\end{axis}
	\end{tikzpicture}
	
	\caption{Sound speed: analytical solution versus numerical measurement with
		varying specific heat ratio. Parameters are $\pr=2.1$, $\tau_{21}=0.501$,
		$\nu_b/\nu=100$, and $\tau_1 = \tau_3$. }
	\label{figure:sound_speed}
\end{figure}
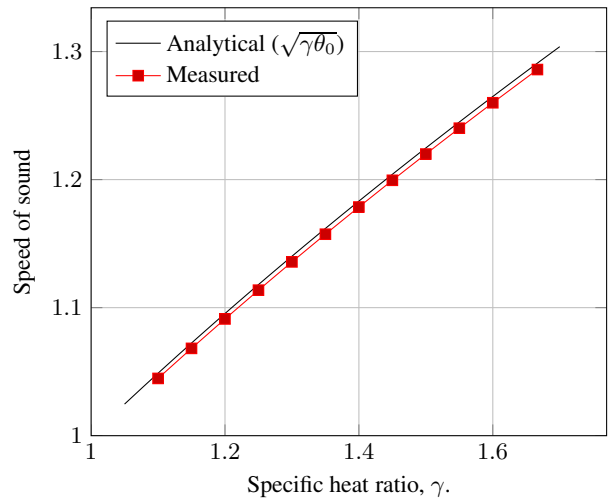

Shown in Fig.~\ref{figure:sound_speed} are the sound speed measured through the
angular frequency of the pressure perturbation. For the usual range of specific
heat ratio, numerical measurements agree well with theoretical predictions.

\section{Conclusions and Discussion}\label{sec:discussion}

In summary, we point out that the collision operator can be defined as a
spectral expansion where the eigen states to which separate relaxation times can
be assigned correspond to the irreducible representations of SO(3) to preserve
rotational symmetry.  For the second moment, two relaxation times are permitted
which give the shear and bulk viscosities.  A kinetic model with arbitrarily
adjustable bulk viscosity is constructed and numerically verified.  An
interesting future direction is that the same decomposition can be applied to
the higher moments.  For instance, according to group theory the 27-dimensional
space of rank-3 tensors can be decomposed into seven lower-dimensional
sub-spaces (sometimes noted as $3\otimes 3\otimes 3=7\oplus 5\oplus 5\oplus
3\oplus 3\oplus 3\oplus 1$) each of which is closed under spatial rotation.
Obviously many of these seven sub-spaces are not fully symmetric as the third
moment should be. It would be interesting to know the maximum number of
relaxation times a given order of moment can accommodate, and physical transport
coefficients the relaxation times correspond to.

\begin{acknowledgments}
This work was supported by the National Science Foundation of China Grants:
No.91741101 and No.91752204, the project of Science and Technology Innovation
Committee of Shenzhen City:K19325001. X.L. acknowledges financial support of
SUSTech Presidential Postdoctoral Fellowship.  An anonymous internet user is
acknowledged for pointing out the rank-3 tensor decomposition result.
\end{acknowledgments}


%

\end{document}